\documentclass[12pt]{article}
\begin{document}
\vspace*{-.6in}
\thispagestyle{empty}
\begin{flushright}
CALT-68-2723\\
\end{flushright}
\baselineskip = 18pt

\vspace{.5in}
{\LARGE
\begin{center}
THE SCIENTIFIC CONTRIBUTIONS\\ OF JO\"EL SCHERK\end{center}}
\vspace{.5in}

\begin{center}
John H. Schwarz
\\
\emph{California Institute of Technology\\ Pasadena, CA  91125, USA}
\end{center}
\vspace{1in}

\begin{center}
\textbf{Abstract}
\end{center}
\begin{quotation}
\noindent Jo\"el Scherk (1946--1980) was an important early
contributor to the development of string theory. Together with
various collaborators, he made numerous profound and influential
contributions to the subject throughout the decade of the 1970s. On
the occasion of a conference at the \'Ecole Normale Sup\'erieure in
2000 that was dedicated to the memory of Jo\"el Scherk I gave a talk
entitled ``Reminiscences of Collaborations with Jo\"el Scherk''
\cite{Schwarz:2000}. The present article, an expanded version of
that one, also discusses work in which I was not involved.
\end{quotation}

\vfil
\centerline{Contribution to a volume entitled {\it The Birth of String Theory}}

\newpage

\pagenumbering{arabic}

\section{Introduction}

Jo\"el Scherk and Andr\'e Neveu were two brilliant French
theoretical physicists who emerged in the latter part of the 1960s.
They were students together at the elite \'Ecole Normale
Sup\'erieure in Paris and in Orsay, where they studied
electromagnetic and final-state interaction corrections to
nonleptonic kaon decays \cite{Neveu:1970tn} under the guidance of
Claude Bouchiat and Philippe Meyer. They defended their ``th\`ese de
troisi\`eme cycle'' (the French equivalent of a PhD) in 1969, and
they were hired together by the CNRS that year \cite{Neveu:2008}.
(Tenure at age 23!) In September 1969 the two of them headed off to
Princeton University. Jo\"el was supported by a NATO
Fellowship.\footnote{President De Gaulle had pulled France out of
the military part of NATO in 1966, but not out of the
educational/scientific/cultural part. See \cite{Neveu:2008} for
an amusing account of how Jo\"el learned that NATO support was
possible.}

In 1969 my duties as an Assistant Professor in Princeton included
advising some assigned graduate students. The first advisees, who
came together to see me, were Andr\'e Neveu and Jo\"el Scherk. I had
no advance warning about them, and so I presumed they were just
another pair of entering students. I certainly didn't know that they
had permanent jobs in France. Because their degrees were not called
PhDs, Princeton University classified them as graduate students, and
so (by the luck of the draw) they were assigned to me. At our first
meeting, I asked the usual questions: ``Do you need to take a course
on electrodynamics?",  ``Do you need to take a course on quantum
mechanics?'', etc. They assured me that they already had learned all
that, so it wouldn't be necessary. I said okay, signed their cards,
and they left.

\section{Loop Amplitudes}

Veneziano discovered his famous formula for a four-particle
amplitude in 1968 \cite{Veneziano:1968}. In 1969 various groups
constructed $N$-particle generalizations of the Veneziano amplitude
\cite{Bardakci:1969a}--\cite{Koba:1969b} and showed that they could
be consistently factorized on a well-defined spectrum of
single-particle states as required for the tree approximation of a
quantum theory \cite{Fubini:1969a}--\cite{Fubini:1970}. In those
days the theory in question was called the {\it dual resonance
model}. Today we would refer to it as the {\it bosonic string
theory}. Knowing the tree-approximation spectrum and couplings, it
became possible to construct one-loop amplitudes. The first such
attempt was made by Kikkawa, Sakita, and Virasoro
\cite{Kikkawa:1969}. They did not have enough information in hand to
do it completely right, but they pioneered many of the key ideas and
pointed the way for their successors. Around this time (the fall of
1969) I began studying these one-loop amplitudes in collaboration
with David Gross, who was also an Assistant Professor at Princeton.

A couple months after our first meeting, Jo\"el and Andr\'e
reappeared in my office and said that they had found some results
they would like to show me. They proceeded to explain their analysis
of the divergence in the planar one-loop amplitude. They had
realized that by performing a Jacobi transformation of the theta
functions in the integrand they could isolate the divergent piece
and propose a fairly natural counterterm \cite{Neveu:1970}.  I was
very impressed by this achievement. It certainly convinced me that
they did not need to take any more quantum mechanics courses!
The modern interpretation of their result is that, viewed in a dual
channel, there is a closed string going into the vacuum. The
divergence can be attributed to the tachyon in that channel, and its
contribution is the piece that they subtracted. This interpretation
explains why such divergences do not occur in theories without
closed-string tachyons.

Since Andr\'e and Jo\"el were working on problems that were very
closely related to those that David and I were studying, we decided
to join forces. Even though we were not yet thinking in terms of
string world sheets, we were able to relate the classification of
dual resonance model amplitudes to the topological classification of
Riemann surfaces with boundaries \cite{Gross:1970a}.
Another discovery by the four of us was that the nonplanar loop
amplitude contains unexpected singularities \cite{Gross:1970b}.
These appeared in addition to the expected two-particle threshold
singularities.\footnote{This was discovered independently by Frye and
Susskind at about the same time \cite{Frye:1970b}.} We assumed, of
course, that the dimension of spacetime is four, since nobody had
yet suggested otherwise. The Virasoro constraints, which should be
taken into account for the internal states circulating in the loop,
had not yet been discovered. As a result, the singularities that we
found were unitarity-violating branch points. We wanted to identify
the leading Regge trajectory associated to these singularities with
the Pomeron, since it carried vacuum quantum numbers, but clearly
something wasn't quite right.

About a year later Lovelace observed that if one chooses the
spacetime dimension to be 26 (one time and 25 space) and supposes
that the Virasoro conditions imply that only transverse excitations
contribute, then instead of branch points the singularities would be
poles \cite{Lovelace:1971}, which could be interpreted as new states
in the spectrum without violating unitarity. As we now know, these
are the closed-string states in the nonplanar open-string loop.
Nowadays this is interpreted as {\it open-string closed-string
duality} of the cylinder diagram. This calculation showed that
unitarity requires choosing the dimension to be 26 and the Regge
intercept value to be one, since these are requirements for the
Virasoro conditions to be satisfied.

Jo\"el left Princeton in the spring of 1970 to spend about six months
in Berkeley. While he was there he collaborated with Michio Kaku,
who was a student of Stanley Mandelstam at the time, studying
divergences in multiloop planar amplitudes \cite{Kaku:1971pe,Kaku:1971pj}.
This was a very ambitious project, given the state of the art in those days.

\section{Orsay, CERN, and NYU}

Scherk's NATO Fellowship was only good for one year, so following his
visits to Princeton and Berkeley, he returned to Orsay for a year followed
by two years at CERN. While he was at Orsay, Jo\"el pioneered the idea of
considering string theories (or dual models) in a zero-slope limit,
which is equivalent to a low-energy limit \cite{Scherk:1971}. In
particular, Jo\"el and Andr\'e studied the massless open-string
spin-one states and showed that in a suitable low-energy limit they
interacted precisely in agreement with Yang--Mills theory
\cite{Neveu:1972}, and they studied the associated gauge invariance
that this implied for the full string theory \cite{Neveu:1972nq}.
These studies made it clear that open strings and their interactions
could be viewed as short-distance modifications of Yang--Mills
theory. This important observation certainly influenced the future evolution
of the subject.

During the two-year period that Jo\"el spent at CERN, beginning in
August 1971, he shared an office with Lars Brink. At first, he
collaborated with Eug\`ene Cremmer on the study of currents and
off-shell amplitudes in string theory as well as factorization in
the closed string (or ``Pomeron'') sector
\cite{Cremmer:1973rn,Cremmer:1972,Cremmer:1973bc}. After that, he
collaborated with Lars Brink, David Olive, and Claudio Rebbi on the
study of scattering amplitudes for Ramond fermions
\cite{Olive:1974sv,Brink:1973jd}.

By now, Jo\"el was very well-known and in much demand. He decided to
accept invitations from NYU and Caltech for the academic year
1973--74, spending the Fall term at NYU and the Winter and Spring
terms at Caltech. During his visit to NYU, he wrote a very elegant
and highly-cited review of string theory for the {\it Reviews of
Modern Physics} \cite{Scherk:1975}. This article was based on a course
that he taught at NYU. He also completed a paper with
Eug\`ene concerning gauge symmetry breaking \cite{Cremmer:1973mg} in
the string theory context.

\section{String Theory for Unification}

In 1972 I left Princeton and moved to Caltech. At Caltech, Murray
Gell-Mann provided funds for me to invite collaborators of my choosing.
One of them was Jo\"el Scherk, who spent the first half of 1974 visiting
Caltech. The timing couldn't have been better.

The hadronic interpretation of string theories was plagued not only
by the occurrence of massless vector particles in the open-string
spectrum, but by a massless tensor particle in the closed-string
spectrum, as well. Several years of effort were expended on trying
to modify each of the two string theories (bosonic and RNS) so as to
lower the leading open-string Regge intercept from 1 to 1/2 and the
leading closed-string Regge intercept from 2 to 1, since these were
the values required for the leading meson and Pomeron Regge
trajectories. Some partial successes were achieved, but no fully
consistent scheme was found. Efforts to modify the critical
spacetime dimension from 26 or 10 to four also led to difficulties.

By 1974, almost everyone who had been working on string theory had
dropped it and moved to greener pastures. The standard model had
been developed, and it was working splendidly. Against this
backdrop, Jo\"el and I stubbornly decided to return to the nagging
unresolved problems of string theory. We felt that string theory has
such a compelling mathematical structure that it ought to be good
for something. Before long our focus shifted to the question of
whether the massless spin two particle in the spectrum interacted at
low energies in accordance with the dictates of general relativity,
so that it might be identified as a graviton. As was mentioned
previously, Jo\"el and Andr\'e Neveu had studied the massless
open-string spin-one states a few years earlier and showed that in a
suitable low-energy limit they interacted precisely in agreement
with Yang--Mills theory \cite{Scherk:1971, Neveu:1972}. Now we
wondered about the analogous question for the massless spin-two
closed-string state. Roughly, what we proved was that critical
string theories have the gauge invariances required to decouple
unphysical polarization states \cite{Scherk:1974a}. Then it followed
on general grounds, which had been elaborated previously by Weinberg
\cite{Weinberg:1965rz}, that the interactions at low energy must be
those of general relativity.

Once we had digested the fact that string theory inevitably contains
gravity we were very excited. We knew that string theory does not
have ultraviolet divergences, because the short-distance structure
is smoothed out, but any field-theoretic approach to gravitation
inevitably gives nonrenormalizable ultraviolet divergences.
Evidently, the way to make a consistent quantum theory of gravity is
to posit that the fundamental entities are strings rather than point
particles \cite{Scherk:1974a}.\footnote{For a more detailed
discussion see \cite{Schwarz2007} in this volume.}

Adopting this viewpoint meant that the fundamental length scale of
string theory, called the string scale, should be identified with
the Planck scale in order to give the correct value for Newton's
constant, at least if one assumes that the size of the extra
dimensions is also given by the string scale. The Planck scale thus
replaces the QCD scale, which was the natural choice for the string
scale when string theory was being developed as a framework for
describing strong interactions (hadron physics). This is a change of
some 19 or 20 orders of magnitude. Even though the mathematics was
largely unchanged, this was a large conceptual leap. Convinced of
the importance of this viewpoint, we submitted a short essay
summarizing the argument to the Gravity Research Foundation's 1975
Essay competition \cite{Scherk:1975a}. The judges were not very
impressed, so we only received an Honorable Mention.

Our 1974 paper proposed changing the goal of string theory to the
problem of constructing a consistent quantum theory of gravity.
Since we already knew from the earlier work of Andr\'e and Jo\"el
that string theory also contains Yang-Mills gauge interactions, it
was natural for us to propose further that string theory should
describe all the other forces at the same time.  This means
interpreting string theory as a unified quantum theory of all
fundamental particles and forces -- an explicit realization of {\it
Einstein's dream}. Moreover, we realized that the existence of extra
dimensions could now be a blessing rather than a curse. After all,
in a gravity theory the geometry of spacetime is determined
dynamically, and one could imagine that this would require, or at
least allow, the extra dimensions to form a very small compact
space. We attempted to construct a specific compactification
scenario in a subsequent paper \cite{Scherk:1975b}. From today's
vantage point, that construction looks rather primitive.

Tamiaki Yoneya independently realized that the massless spin two
state of string theory interacts at low energy in accordance with
the dictates of general relativity \cite{Yoneya:1974}. Indeed, his
paper appeared first, though we were not aware of it at the time.
However, as Yoneya graciously acknowledges, Jo\"el and I were the
only ones to take the next step and to propose that string theory
should be the basis for constructing a unified quantum theory of all
forces. The recognition of that possibility represented a turning
point in my research career. I found the case compelling, and I
became committed to exploring its implications. I believe that
Jo\"el felt the same way. I still do not understand why it took
another decade until a large segment of the theoretical physics
world became convinced that string theory was the right approach to
unification. (There were some people who caught on earlier, of
course.) One of my greatest regrets is that Jo\"el was not able to
witness the impact that this idea eventually would have.

During Jo\"el's Caltech visit, we also explored some aspects of
gravity that could be affected by the string interpretation. One
paper interpreted the three-form flux $H=dB$ in terms of spacetime
torsion \cite{Scherk:1974mc}. Since we knew that the light-cone
gauge is convenient for exploring certain aspects of string theory,
we also attempted to formulate general relativity in the light-cone
gauge \cite{Scherk:1974zm}.

After leaving Caltech, Jo\"el participated in a
summer 1974 workshop on string theory
at the Aspen Center for Physics, which I organized. This had been
planned a year earlier, when there were still quite a few people working in
string theory. My memory is fading, but I do not recall the participants
showing much interest in our proposal to use string theory for
unification.

After Aspen, Jo\"el returned to Paris, since his group at Orsay had
moved from Orsay to the Laboratoire de Physique Th\'eorique of the
\'Ecole Normale Sup\'erieure in Paris. Aside from a few months in
Cambridge in 1977 and brief visits elsewhere, this is where Jo\"el
spent the remainder of his career. At the LPTENS, Jo\"el resumed his
collaboration with Eug\`ene Cremmer. Soon, they turned out a pair of
well-known papers \cite{Cremmer:1975sj,Cremmer:1976ir} that grappled
with issues raised by the unification interpretation of string
theory. The first paper \cite{Cremmer:1975sj} introduced the notion
of {\em winding numbers} for the first time. As is now well-known,
closed strings can wrap on cycles in the compact dimensions. This
possibility was important in the later construction of the heterotic
string as well as for the discovery of T-duality almost a decade
later.\footnote{Eug\`ene informs me that they failed to discover
T-duality because of their field theory prejudices. This shows that
even dedicated string theorists had such prejudices. 
See \cite{Cremmer:2008} for further discussion of this.} The second
paper \cite{Cremmer:1976ir} emphasized the idea that the compact
internal spaces cannot be chosen arbitrarily; instead, they are
fixed by the mechanism that they called ``spontaneous
compactification''. This means that they must be stable or
metastable solutions of the equations of
motion.\footnote{Metastability implies massless moduli, which are
ruled out experimentally by tests of the tensor nature of the
gravitational force.}

\section{Spacetime Supersymmetry}

The super-Virasoro gauge symmetry of the Ramond--Neveu--Schwarz
model \cite{Ramond:1971b,Neveu:1971b} describes the superconformal
symmetry of the two-dimensional world-sheet theory. The
supersymmetry of the two-dimensional world-sheet action was
described later in 1971 by Gervais and Sakita \cite{Gervais:1971b}.
This was the first example of a supersymmetric quantum field theory.
For about five years, the supersymmetry considered by string
theorists only pertained to the two-dimensional world-sheet theory.
It did not occur to us that there could also be supersymmetry in
ten-dimensional spacetime.

Bruno Zumino, who was also at CERN when Jo\"el was there, became
very interested in the RNS string's gauge conditions associated to
the two-dimensional superconformal algebra and discussed it at
length with Jo\"el and Lars. His work on this subject is described
in \cite{Zumino:1974}. Following that, he and Julius Wess began to
consider the possibility of constructing four-dimensional field
theories with analogous features. This resulted in their famous work
\cite{Wess:1974tw} on globally supersymmetric field theories in four
dimensions. As a consequence of their paper,\footnote{The work of
Golfand and Likhtman \cite{Golfand:1971iw}, which was the first to
introduce the four-dimensional super-Poincar\'e group, was not known
in the West at that time.} supersymmetry quickly became an active
research topic. A few years later came the discovery of supergravity
theories \cite{Freedman:1976xh,Deser:1976eh}.

Lars Brink, Jo\"el, and I constructed supersymmetric Yang-Mills
theories in various dimensions \cite{Brink:1977}. When this work was
carried out, Lars and I were at Caltech and Jo\"el was in Paris. (We
communicated by mail, since email was not yet an option.) We
discovered that the requisite gamma-matrix identity required by
these theories, $\gamma^m_{(ab}\gamma^m_{ c)d} =0$, is satisfied in
dimensions 3, 4, 6, and 10. Dimensional reduction was also
discussed, giving (among other things) the first construction of
${\cal N}=4$ super Yang--Mills theory in four dimensions.
At the same time as the work described above, Jo\"el was
collaborating with Fernando Gliozzi and David Olive on some closely
related ideas \cite{Gliozzi:1976, Gliozzi:1977}. This threesome is
now referred to as GSO. (Lars and I were not aware of this collaboration
until we saw their papers.) The GSO papers also explored super Yang--Mills
theories. Moreover, they took the next major step, which concerned
the RNS string theory.

GSO proposed a projection of the RNS spectrum -- {\em the GSO
Projection} -- that removes roughly half of the states (including
the tachyon). Specifically, in the bosonic (NS) sector they
projected away the odd G-parity states, a possibility that was
discussed earlier, and in the fermionic (R) sector they projected
away half the states, keeping only certain definite chiralities.
Then they counted the remaining physical degrees of freedom
of the free string at each
mass level. After the GSO projection the masses of open-string
states, for both bosons and fermions, are given by $\alpha' M^2 =n$,
where $n=0,1,2,\ldots$ Denoting the open-string degeneracies of
states in the GSO-projected theory by $d_{\rm NS}(n)$ and $d_{\rm
R}(n)$, they showed that these are encoded in the generating
functions
$$
f_{\rm NS}(w) = \sum_{n=0}^\infty d_{\rm NS}(n) w^n$$ $$=
\frac{1}{2\sqrt{w}} \left[\prod_{m=1}^{\infty}
\left(\frac{1+w^{m-1/2}}{1-w^m}\right)^8 - \prod_{m=1}^{\infty}
\left(\frac{1 -w^{m-1/2}}{1-w^m}\right)^8 \right].
$$
and
$$
f_{\rm R}(w) = \sum_{n=0}^\infty d_{\rm R}(n) w^n = 8
\prod_{m=1}^{\infty} \left(\frac{1+w^{m}}{1-w^m}\right)^8 .
$$

In 1829, Jacobi proved the remarkable identity \cite{Jacobi:1829}
$$
f_{\rm NS}(w) = f_{\rm R}(w),
$$
though he used a different notation.\footnote{Jacobi's paper
acknowledges an assistant named Scherk!} Thus, there are an equal
number of bosons and fermions at every mass level, as required. This
was compelling evidence (though not a proof) for {\em
ten-dimensional spacetime supersymmetry} of the GSO-projected
theory. Prior to this work, one knew that the RNS theory has
world-sheet supersymmetry, but the realization that the theory
should have spacetime supersymmetry was a major advance. I was very
delighted by this result. One could now envisage a tachyon-free
string theory that would make sense as a starting point for a
unified theory.

At about the same time as the GSO breakthrough, Jo\"el collaborated
with Sergio Ferrara, Dan Freedman, Peter Van Nieuwenhuizen, and
Bruno Zumino on various different topics in supergravity
\cite{Ferrara:1976kg}--\cite{Ferrara:1976iq}. In the following year
(1977) he continued studying supergravity, now collaborating mostly
with Eug\`ene \cite{Cremmer:1977zt}--\cite{Cremmer:1978bh}. Two of
these papers \cite{Cremmer:1977tc,Cremmer:1977za} were written while
Jo\"el spent several months in the Spring of 1977 at the DAMTP in
Cambridge. Eug\`ene and Jo\"el communicated by snail mail and FAX
during that period.

Eug\`ene, Sergio, and Jo\"el reformulated ${\cal N} =4$ supergravity 
in a manifestly $SU(4)$-invariant form that was motivated by $N=1$ 
supergravity in ten dimensions, which was itself motivated by superstring 
theory. It exhibited an on-shell $SU(1,1)$ duality invariance
\cite{Cremmer:1977tt}. This was the first discovery of the duality
symmetries of extended supergravity theories. S-duality in string
theory is such a duality, as was understood some 17 years later.
The discovery of such dualities was very influential in the construction
of ${\cal N} =8$ supergravity by Eug\`ene and Bernard Julia \cite{Cremmer:1978ds,Cremmer:1979up}. The duality group in that case
is $E_{7,7}$. Eug\`ene informs me that health problems prevented Jo\"el
from participating in that collaboration.

One of the most beautiful results in supergravity, and Jo\"el's
most-cited paper, is the March 1978 construction of the action of
11-dimensional supergravity by Eug\`ene, Bernard, and Jo\"el
\cite{Cremmer:1978}.
It was immediately clear that 11-dimensional supergravity is very
beautiful, and it aroused a lot of interest. However, it was
puzzling for a long time how it fits into the greater scheme of
things and whether it has any connection to string theory. Clearly,
supergravity in 11 dimensions is not a consistent quantum theory by
itself, since it is very singular in the ultraviolet. Moreover,
since superstring theory only has ten dimensions, it did not seem
possible that it could serve as a regulator. It took more than
fifteen years to find the answer to this conundrum
\cite{Townsend:1995kk,Witten:1995ex}: At strong coupling Type IIA
superstring theory develops a circular 11th dimension whose radius
grows with the string coupling constant. In the limit of infinite
coupling one obtains {\it M theory}, which is presumably a
well-defined quantum theory that has 11 noncompact dimensions.
Eleven-dimensional supergravity is the leading low-energy
approximation to M theory. In other words, M theory is the UV
completion of 11-dimensional supergravity.

Later in 1978 Eug\`ene, Bernard, and Jo\"el teamed up with Sergio
Ferrara, Luciano Girardello, and Peter Van Nieuwenhuizen on a pair
of papers studying the super Higgs effect in four-dimensional
supergravity theories coupled to matter
\cite{Cremmer:1978iv,Cremmer:1978hn}. This work has been used a
great deal in subsequent studies.

\section{Supersymmetry Breaking}

I spent the academic year 1978--79 visiting the LPTENS, on
leave of absence from Caltech. I was eager to work with Jo\"el on
supergravity, supersymmetrical strings, and related matters. He was
struggling with rather serious health problems during that year, so
he wasn't able to participate as fully as when we were in Caltech
five years earlier, but he was able to work about half time. On that
basis we were able to collaborate successfully.

After various wide-ranging discussions we decided to focus on the
problem of supersymmetry breaking. We wondered how, starting from a
supersymmetric string theory in ten dimensions, one could end up
with a nonsupersymmetric world in four dimensions. The specific
supersymmetry breaking mechanism that we discovered can be explained
classically and does not really require strings, so we explored it
in a field theoretic setting \cite{Scherk:1979a,Scherk:1979b}.
The idea is that in a theory with
extra dimensions and global symmetries that do not commute with
supersymmetry ($R$ symmetries and $(-1)^F$ are examples), one could
arrange for a twisted compactification, and that this would break
supersymmetry. For example, if one extra dimension forms a circle,
the fields when continued around the circle could come could back
transformed by an R-symmetry group element. If the gravitino, in
particular, is transformed then it acquires mass in a consistent
manner.

An interesting example of our supersymmetry breaking mechanism was
worked out in a paper we wrote together with Eug\`ene
\cite{Cremmer:1979}. We started with maximal supergravity in five
dimensions. This theory contains eight gravitinos that transform in
the fundamental representation of a USp(8) R-symmetry group. We took
one dimension to form a circle and examined the resulting
four-dimensional theory keeping the lowest Kaluza--Klein modes. The
supersymmetry-breaking R-symmetry element is a USp(8) element that
is characterized by four real mass parameters, since this group has
rank four. These four masses give the masses of the four complex
gravitinos of the resulting four-dimensional theory. In this way we
were able to find a consistent four-parameter deformation of ${\cal
N} = 8$ supergravity.

Even though the work that Jo\"el and I did on supersymmetry breaking
was motivated by string theory, we only discussed field theory
applications in our articles. The reason I never wrote about string
theory applications was that in the string theory setting it did not
seem possible to decouple the supersymmetry breaking mass parameters
from the compactification scales.  This was viewed as a serious
problem, because the two scales are supposed to be hierarchically
different. In recent times, people have been considering string
theory brane-world scenarios in which much larger compactification
scales are considered. In such a context our supersymmetry breaking
mechanism might have a role to play. Indeed, quite a few authors
have explored various such possibilities.

\section{Concluding Comments}

When I left Paris in the summer of 1979 I visited CERN for a month.
There I began a collaboration with Michael Green. During that month
we began to formulate a plan for exploring how the spacetime
supersymmetry identified by GSO is realized in the interacting
string theory. In September 1979, when I spoke at a conference on
supergravity that was held in Stony Brook, we did not yet have
definitive results. Therefore, I reported on the work that Jo\"el
and I had done on supersymmetry breaking \cite{Schwarz:1979}. Jo\"el
gave a talk entitled {\it From Supergravity to Antigravity} at the
Stony Brook conference \cite{Scherk:1979aj,Scherk:1979c}. He was
intrigued by the fact that graviton exchanges in string theory are
accompanied by antisymmetric tensor and scalar exchanges that can
cancel the gravitational attraction. Nowadays we understand that the
effect that he was discussing is quite important. For example,
parallel BPS D-branes form stable supersymmetric systems precisely
because the various forces cancel. The Stony Brook conference was
the last time that I saw Jo\"el.

In March 1980 Jo\"el attended a meeting in Erice, Sicily. Lars Brink, who
was also there, recalls being very worried about Jo\"el's health.
Six weeks after that meeting he passed away, which came as a great shock
to his many friends and colleagues. The ideas that Jo\"el
pioneered during the decade of the 1970s have been very influential in
the subsequent decades. It is a pity that he could not
participate in these developments and enjoy the recognition that
he would have received.

\section*{Acknowledgments}

I am grateful to Lars Brink, Eug\`ene Cremmer, Andr\'e Neveu, and
David Olive for reading this manuscript and making numerous helpful
suggestions. This work was supported in part by the U.S. Dept. of
Energy under Grant No. DE-FG03-92-ER40701.


\newpage

\begin{thebibliography}{99}

\bibitem{Schwarz:2000}
J.~H. Schwarz, ``Reminiscences of Collaborations with Jo\"el Scherk,''
arXiv:hep-th/0007117.

\bibitem{Neveu:1970tn}
A.~Neveu and J.~Scherk, ``Final-State Interaction and Current
Algebra in $K \to 3\pi$ and $\eta\to 3\pi$ Decays,'' Annals Phys.\
{\bf 57} (1970) 39.

\bibitem{Neveu:2008}
A.~Neveu, ``Personal Recollections about the First Three Years of
String Theory,'' this volume.

\bibitem{Veneziano:1968}
G.~Veneziano, ``Construction of a Crossing-Symmetric Regge-Behaved
Amplitude for Linearly Rising Regge Trajectories,'' Nuovo Cim. {\bf
57A} (1968) 190.

\bibitem{Bardakci:1969a}
K.~Bardakci and H.~Ruegg, ``Reggeized Resonance Model for Arbitrary
Production Processes,'' Phys. Rev. {\bf 181} (1969) 1884.

\bibitem{Goebel:1969}
C.~J.~Goebel and B.~Sakita, ``Extension of the Veneziano Formula to
$N$-Particle Amplitudes,'' Phys. Rev. Lett. {\bf 22} (1969) 257.

\bibitem{Chan:1969b}
H.~M.~Chan and T.~S.~Tsun, ``Explicit Construction of the $N$-Point
Function in the Generalized Veneziano Model,'' Phys. Lett. {\bf 28B}
(1969) 485.

\bibitem{Koba:1969a}
Z.~Koba and H.~B.~Nielsen, ``Reaction Amplitudes for $N$ Mesons, a
Generalization of the Veneziano--Bardakci--Ruegg--Virasoro Model,''
Nucl. Phys. {\bf B10} (1969) 633.

\bibitem{Koba:1969b}
Z.~Koba and H.~B.~Nielsen, ``Manifestly Crossing-Invariant
Parametrization of the $N$-Meson Amplitude,'' Nucl. Phys. {\bf B12}
(1969) 517.

\bibitem{Fubini:1969a}
S.~Fubini and G.~Veneziano, ``Level Structure of Dual Resonance
Models,'' Nuovo Cim. {\bf 64A} (1969) 811.

\bibitem{Fubini:1969b}
S.~Fubini, D.~Gordon, and G.~Veneziano, ``A General Treatment of
Factorization in Dual Resonance Models,'' Phys. Lett. {\bf 29B}
(1969) 679.

\bibitem{Bardakci:1969b}
K.~Bardakci and S.~Mandelstam, ``Analytic Solution of the
Linear-Trajectory Bootstrap," Phys. Rev. {\bf 184} (1969) 1640.

\bibitem{Fubini:1970}
S.~Fubini and G.~Veneziano, ``Duality in Operator Formalism,'' Nuovo
Cim. {\bf 67A} (1970) 29.

\bibitem{Kikkawa:1969}
K.~Kikkawa, B.~Sakita and M.~A.~Virasoro, ``Feynman-like diagrams
compatible with duality. I: Planar diagrams,'' Phys.\ Rev.\  {\bf
184} (1969) 1701.

\bibitem{Neveu:1970}
A.~Neveu and J.~Scherk, ``Parameter-Free Regularization of One-Loop
Unitary Dual Diagram,'' Phys. Rev. {\bf D1} (1970) 2355.


\bibitem{Gross:1970a}
D.~J.~Gross, A.~Neveu, J.~Scherk, and J.~H.~Schwarz, ``The Primitive
Graphs of Dual Resonance Models,'' Phys.\ Lett.\  {\bf 31B} (1970)
592.

\bibitem{Gross:1970b}
D.~J.~Gross, A.~Neveu, J.~Scherk, and J.~H.~Schwarz,
``Renormalization and Unitarity in the Dual Resonance Model,''
Phys.\ Rev.\  {\bf D2} (1970) 697.

\bibitem{Frye:1970b}
G.~Frye and L.~Susskind, ``Non-Planar Dual Symmetric Loop Graphs and
the Pomeron,'' Phys. Lett. {\bf 31B} (1970) 589.

\bibitem{Lovelace:1971}
C.~Lovelace, ``Pomeron Form Factors and Dual Regge Cuts,'' Phys.
Lett. {\bf 34B} (1971) 500.

\bibitem{Kaku:1971pe}
M.~Kaku and J.~Scherk, ``Divergence of the Two-Loop Planar Graph
in the Dual-Resonance Model,'' Phys.\ Rev.\  D {\bf 3} (1971) 430.

\bibitem{Kaku:1971pj}
M.~Kaku and J.~Scherk, ``Divergence of the N-Loop Planar Graph
in the Dual-Resonance Model,'' Phys.\ Rev.\  D {\bf 3} (1971) 2000.

\bibitem{Scherk:1971}
J.~Scherk, ``Zero Slope Limit of the Dual Resonance Model,'' Nucl.\
Phys.\  {\bf B31} (1971) 222.

\bibitem{Neveu:1972}
A.~Neveu and J.~Scherk, ``Connection Between Yang--Mills Fields and
Dual Models,'' Nucl.\ Phys.\  {\bf B36} (1972) 155.

\bibitem{Neveu:1972nq}
A.~Neveu and J.~Scherk, ``Gauge Invariance and Uniqueness of the
Renormalisation of Dual Models with Unit Intercept,''
Nucl.\ Phys.\  B {\bf 36} (1972) 317.

\bibitem{Cremmer:1973rn}
E.~Cremmer and J.~Scherk, ``Currents and Green's Functions
for Dual Models. (II) Off-Shell Dual Amplitudes,''
Nucl.\ Phys.\  B {\bf 48} (1972) 29.

\bibitem{Cremmer:1972}
E.~Cremmer and J.~Scherk, ``Factorization of the Pomeron Sector and
Currents in the Dual Resonance Model,'' Nucl.\ Phys.\  {\bf B50}
(1972) 222.

\bibitem{Cremmer:1973bc}
E.~Cremmer and J.~Scherk, ``Regge Limit and Scaling in a Dual Model
of Currents,'' Nucl.\ Phys.\  B {\bf 58} (1973) 557.

\bibitem{Brink:1973nb}
L.~Brink, D.~I.~Olive and J.~Scherk, ``The Gauge Properties of the
Dual Model Pomeron-Reggeon Vertex - Their Derivation and Their
Consequences,'' Nucl.\ Phys.\  B {\bf 61} (1973) 173.

\bibitem{Olive:1974sv}
D.~I.~Olive and J.~Scherk, ``Towards Satisfactory Scattering Amplitudes
for Dual Fermions,'' Nucl.\ Phys.\  B {\bf 64} (1973) 334.

\bibitem{Brink:1973jd}
L.~Brink, D.~I.~Olive, C.~Rebbi and J.~Scherk, ``The Missing Gauge
Conditions for the Dual Fermion Emission Vertex and Their
Consequences,'' Phys.\ Lett.\  B {\bf 45} (1973) 379.

\bibitem{Scherk:1975}
J.~Scherk, ``An Introduction to the Theory of Dual Models and
Strings,'' Rev. Mod. Phys. {\bf 47} (1975) 123.

\bibitem{Cremmer:1973mg}
E.~Cremmer and J.~Scherk, ``Spontaneous Dynamical Breaking of Gauge
Symmetry in Dual Models,'' Nucl.\ Phys.\  B {\bf 72} (1974) 117.

\bibitem{Schwarz2007}
J.~H.~Schwarz, The Early Years of String Theory: A Personal
Perspective,'' in this volume.

\bibitem{Scherk:1974a}
J.~Scherk and J.~H.~Schwarz, ``Dual Models for Non-Hadrons,'' Nucl.
Phys. {\bf B81} (1974) 118.

\bibitem{Weinberg:1965rz}
S.~Weinberg, ``Photons and Gravitons in Perturbation Theory:
Derivation of Maxwell's and Einstein's Equations,''
Phys.\ Rev.\  {\bf 138} (1965) B988.

\bibitem{Scherk:1974b}
J.~Scherk and J.~H.~Schwarz, ``Dual Models and the Geometry of
Space-Time,'' Phys. Lett. {\bf B52} (1974) 347.

\bibitem{Scherk:1975a}
J.~Scherk and J.~H.~Schwarz, ``Dual Model Approach to a
Renormalizable Theory of Gravitation,'' Submitted to the 1975
Gravitation Essay Contest of the Gravity Research Foundation.
Reprinted in {\it Superstrings, Vol. 1}, ed. J. Schwarz, World
Scientific (1985).

\bibitem{Scherk:1975b}
J.~Scherk and J.~H.~Schwarz, ``Dual Field Theory of Quarks and
Gluons,'' Phys. Lett. {\bf 57B} (1975) 463.

\bibitem{Yoneya:1974}
T.~Yoneya, ``Connection of Dual Models to Electrodynamics and
Gravidynamics,'' Prog. Theor. Phys. {\bf 51} (1974) 1907.

\bibitem{Scherk:1974mc}
J.~Scherk and J.~H.~Schwarz, ``Dual Models and the Geometry of Space-Time,''
Phys.\ Lett.\  B {\bf 52} (1974) 347.

\bibitem{Scherk:1974zm}
J.~Scherk and J.~H.~Schwarz, ``Gravitation in the Light-Cone Gauge,''
Gen.\ Rel.\ Grav.\  {\bf 6} (1975) 537.

\bibitem{Cremmer:1975sj}
E.~Cremmer and J.~Scherk, ``Dual Models in Four Dimensions with
Internal Symmetries,'' Nucl.\ Phys.\  B {\bf 103} (1976) 399.

\bibitem{Cremmer:1976ir}
E.~Cremmer and J.~Scherk, ``Spontaneous Compactification of Space in
an Einstein Yang-Mills Higgs Model,'' Nucl.\ Phys.\  B {\bf 108} (1976) 409.

\bibitem{Cremmer:2008}
E.~Cremmer, ``Person Recollections about the Birth of String Theory,''
this volume.

\bibitem{Ramond:1971b}
P.~Ramond, ``Dual Theory for Free Fermions,'' Phys. Rev, D {\bf 3}
(1971) 2415.

\bibitem{Neveu:1971b}
A.~Neveu and J.~H.~Schwarz, ``Factorizable Dual Model of Pions,''
Nucl. Phys. B {\bf 31} (1971) 86.

\bibitem{Gervais:1971b}
J.~L.~Gervais and B.~Sakita, ``Field Theory Interpretation of
Supergauges in Dual Models,'' Nucl. Phys. B {\bf 34} (1971) 632.

\bibitem{Zumino:1974}
B.~Zumino, ``Relativistic Strings and Supergauges,'' p. 367 in {\it
Renormalization and Invariance in Quantum Field Theory,} ed. E.
Caianiello (Plenum Press, 1974).

\bibitem{Wess:1974tw}
J.~Wess and B.~Zumino, ``Supergauge Transformations in Four-Dimensions,''
Nucl.\ Phys.\  B {\bf 70} (1974) 39.

\bibitem{Golfand:1971iw}
Yu.~A.~Golfand and E.~P.~Likhtman, ``Extension of the Algebra of
Poincar\'e Group Generators and Violation of P Invariance,'' {\it
JETP Lett.}\  {\bf 13} (1971) 323 [{\it Pisma Zh.\ Eksp.\ Teor.\
Fiz.}\ {\bf 13} (1971) 452].

\bibitem{Freedman:1976xh}
D.~Z.~Freedman, P.~van Nieuwenhuizen and S.~Ferrara,
``Progress Toward a Theory of Supergravity,''
Phys.\ Rev.\  D {\bf 13} (1976) 3214.

\bibitem{Deser:1976eh}
S.~Deser and B.~Zumino, ``Consistent Supergravity,''
Phys.\ Lett.\  B {\bf 62} (1976) 335.

\bibitem{Ferrara:1976um}
  S.~Ferrara, J.~Scherk and P.~van Nieuwenhuizen,
  Phys.\ Rev.\ Lett.\  {\bf 37} (1976) 1035.

\bibitem{Brink:1977}
L. Brink, J. H. Schwarz,  and J. Scherk ,
``Supersymmetric Yang-Mills Theories,''
Nucl. Phys. {\bf B121}, 77 (1977).

\bibitem{Gliozzi:1976}
F.~Gliozzi, J.~Scherk, and D.~Olive,
`Supergravity and the Spinor Dual Model,''
Phys. Lett. {\bf 65B}, 282 (1976).

\bibitem{Gliozzi:1977}
F.~Gliozzi, J.~Scherk, and D.~Olive,
``Supersymmetry, Supergravity Theories and the Dual Spinor Model,''
Nucl. Phys. {\bf B122}, 253 (1977).

\bibitem{Ferrara:1976kg}
S.~Ferrara, D.~Z.~Freedman, P.~van Nieuwenhuizen, P.~Breitenlohner,
F.~Gliozzi and J.~Scherk, ``Scalar Multiplet Coupled to Supergravity,''
Phys.\ Rev.\  D {\bf 15} (1977) 1013.

\bibitem{Jacobi:1829}
C.~G.~J. Jacobi, {\it Fundamenta}, Konigsberg, 1829.

\bibitem{Ferrara:1976ni}
S.~Ferrara, F.~Gliozzi, J.~Scherk and P.~Van Nieuwenhuizen,
``Matter Couplings in Supergravity Theory,''
Nucl.\ Phys.\  B {\bf 117} (1976) 333.

\bibitem{Ferrara:1976vf}
S.~Ferrara, J.~Scherk and B.~Zumino,
``Supergravity and Local Extended Supersymmetry,''
Phys.\ Lett.\  B {\bf 66} (1977) 35.

\bibitem{Ferrara:1976iq}
S.~Ferrara, J.~Scherk and B.~Zumino,
``Algebraic Properties of Extended Supergravity Theories,''
Nucl.\ Phys.\  B {\bf 121} (1977) 393.

\bibitem{Cremmer:1977zt}
E.~Cremmer, J.~Scherk and S.~Ferrara,
``U(N) Invariance in Extended Supergravity,''
Phys.\ Lett.\  B {\bf 68} (1977) 234.

\bibitem{Horvath:1977st}
Z.~Horvath, L.~Palla, E.~Cremmer and J.~Scherk,
``Grand Unified Schemes and Spontaneous Compactification,''
Nucl.\ Phys.\  B {\bf 127} (1977) 57.

\bibitem{Cremmer:1977tc}
E.~Cremmer and J.~Scherk, ``Algebraic Simplifications in
Supergravity Theories,'' Nucl.\ Phys.\  B {\bf 127} (1977) 259.

\bibitem{Cremmer:1977za}
E.~Cremmer and J.~Scherk, ``Modified Interaction of the Scalar
Multiplet Coupled to Supergravity,'' Phys.\ Lett.\  B {\bf 69}
(1977) 97.

\bibitem{Cremmer:1977tt}
E.~Cremmer, J.~Scherk and S.~Ferrara,
``SU(4) Invariant Supergravity Theory,''
Phys.\ Lett.\  B {\bf 74} (1978) 61.

\bibitem{Cremmer:1978bh}
E.~Cremmer and J.~Scherk,
``The Supersymmetric Nonlinear Sigma Model in Four-Dimensions and Its
Coupling to Supergravity,'' Phys.\ Lett.\  B {\bf 74} (1978) 341.

\bibitem{Cremmer:1978}
E.~Cremmer, B.~Julia, and J.~Scherk, ``Supergravity Theory in 11
Dimensions,'' Phys. Lett. {\bf76B} (1978) 409.

\bibitem{Cremmer:1978ds}
E.~Cremmer and B.~Julia, ``The N=8 Supergravity Theory. 1.
The Lagrangian,'' Phys.\ Lett.\  B {\bf 80} (1978) 48.

\bibitem{Cremmer:1979up}
E.~Cremmer and B.~Julia, ``The SO(8) Supergravity,''
Nucl.\ Phys.\  B {\bf 159} (1979) 141.

\bibitem{Townsend:1995kk}
P.~K.~Townsend, ``The Eleven-Dimensional Supermembrane Revisited,''
Phys.\ Lett.\  B {\bf 350} (1995) 184 [arXiv:hep-th/9501068].

\bibitem{Witten:1995ex}
E.~Witten, ``String Theory Dynamics in Various Dimensions,''
Nucl.\ Phys.\  B {\bf 443} (1995) 85 [arXiv:hep-th/9503124].

\bibitem{Cremmer:1978iv}
E.~Cremmer, B.~Julia, J.~Scherk, P.~van Nieuwenhuizen, S.~Ferrara and L.~Girardello,
``Superhiggs Effect in Supergravity with General Scalar Interactions,''
Phys.\ Lett.\  B {\bf 79} (1978) 231.

\bibitem{Cremmer:1978hn}
E.~Cremmer, B.~Julia, J.~Scherk, S.~Ferrara, L.~Girardello and P.~van Nieuwenhuizen,
``Spontaneous Symmetry Breaking and Higgs Effect in Supergravity Without
Cosmological Constant,'' Nucl.\ Phys.\  B {\bf 147} (1979) 105.

\bibitem{Scherk:1979a}
J.~Scherk and J.~H.~Schwarz,
``Spontaneous Breaking of Supersymmetry Through Dimensional Reduction,''
Phys. Lett. {\bf B82}, 60 (1979).

\bibitem{Scherk:1979b}
J.~Scherk and J.~H.~Schwarz,
``How to Get Masses From Extra Dimensions,''
Nucl. Phys. {\bf B153}, 61 (1979).

\bibitem{Cremmer:1979}
E. Cremmer, J.~Scherk and J.~H.~Schwarz,
``Spontaneously Broken N = 8 Supergravity,''
Phys. Lett. {\bf B84}, 83 (1979).

\bibitem{Schwarz:1979}
J.~H.~Schwarz, ``How to Break Supersymmetry,'' p. 19 in {\it
Supergravity}, eds. P. van Nieuwenhuizen and D.~Z.~Freedman
(North-Holland 1979).

\bibitem{Scherk:1979aj}
J.~Scherk, ``Antigravity: A Crazy Idea?,''
Phys.\ Lett.\  B {\bf 88} (1979) 265.

\bibitem{Scherk:1979c}
J.~Scherk,
``From Supergravity to Antigravity,''
p. 43 in {\it Supergravity}, eds. P. van Nieuwenhuizen and D.~Z.~Freedman
(North-Holland 1979).


\end{thebibliography}
\end{document}